\documentclass[10pt,journal]{IEEEtran}
\usepackage[cmex10]{amsmath}
\usepackage{amssymb}
\usepackage[pdftex]{graphicx}

\usepackage[tight,footnotesize]{subfigure}

\usepackage{url}

\usepackage{siunitx}

\usepackage{cite}

\newcommand{\prt}{p(r,t|r_0)}
\newcommand{\prtzero}{p(r,t \rightarrow 0 |r_0)}

\newcommand{\rrn}{r^{\text{rx}}}
\newcommand{\rtn}{r^{\text{tx}}}

\newcommand{\nrxTsph}[2]{N^{\text{s-tx}}_{\text{s-rx, }#1}(#2)}
\newcommand{\nrxTpnt}[2]{N^{\text{p-tx}}_{\text{s-rx, }#1}(#2)}

\newcommand{\directivityG}[1]{G_{D,\text{omni}}(#1)}
\newcommand{\peakAt}[1]{t_{\text{peak}}(#1)}

\newcommand{\tend}{t_{\text{end}}}

\begin{document}
\title{Chemical Propagation Pattern for Molecular Communications}

\author{H.~Birkan Yilmaz,~\IEEEmembership{Member,~IEEE,}
		Gee-Yong Suk,~\IEEEmembership{Student Member,~IEEE,}
        and\\ Chan-Byoung~Chae,~\IEEEmembership{Senior Member,~IEEE}% <-this % stops a space
%\thanks{Manuscript received Sep. 2, 2016}%; revised \today}% <-this % stops a space
\thanks{This research was in part supported by the MSIP, under the ``ICT Consilience Creative Program" (IITP-R0346-16-1008) and  by the Basic Science Research Program (2014R1A1A1002186), through the NRF of Korea.}% <-this % stops a space
\thanks{H. B.~Yilmaz, G.-Y.~Suk, and C.-B.~Chae are with the School of Integrated Technology, Yonsei Institute of Convergence Technology, Yonsei University, Korea. e-mails: \{birkan.yilmaz, gysuk, cbchae\}@yonsei.ac.kr}}

\markboth{}%
{Chemical Propagation Pattern for Molecular Communications}

\maketitle

\begin{abstract}
In a diffusion-based molecular communication system, molecules are employed to convey information. When propagation and reception processes are considered in a framework of first passage processes, we need to focus on absorbing receivers. For this kind of molecular communication system, the characteristics of the channel is also affected by the shape of the transmitter. In the literature, most studies focus on systems with a point transmitter due to circular symmetry. In this letter, we address propagation and reception pattern for chemical signals emitted from a spherical transmitter. We also investigate the directivity gain achieved by the reflecting spherical transmitter. We quantify the power gain by measuring the received power at different angles on a circular region. Moreover, we define three metrics, i.e., the half-power pattern-width, the directivity gain, and the peak time of the signal, for analyzing the received signal pattern.
\end{abstract}


%\IEEEpeerreviewmaketitle

% % % % % % % % % % % % % % % % % % % % % % % % % % % % % % % %
% % % % % % % % % % % % % % % % % % % % % % % % % % % % % % % %
\section{Introduction}
\label{sec_introduction}

\IEEEPARstart{T}{he} developments in nanotechnology research have resulted in implementation of simple nano and micro nodes capable of carrying out simple tasks. Designing and enabling communication between these nodes emerged as a new need for achieving complex tasks at small scales~\cite{craighead2006futureLO,nakano2013molecularC_book,akyildiz2011nanonetworksAN,Farsad2016ComSurv}. Various molecular communication systems were proposed in the literature, such as molecular communication via diffusion (MCvD), ion signaling, active transport, and bacterium-based communication. Among these systems, a particularly effective and energy-efficient method of exchanging information is MCvD, a short-to-medium range molecular communication technique in which the molecules diffuse in the propagation medium to transfer the intended information. MCvD consists of three main processes: emission, propagation, and absorption. In MCvD, analyzing the received molecule distribution with respect to time is crucial for characterizing the channel~\cite{yilmaz20143dChannelCF}.

Channel characteristics for molecular communication with an absorbing receiver in a 3-dimensional (3D) environment are analyzed in \cite{yilmaz20143dChannelCF}. In \cite{akkaya2015effectOR}, the authors formulated the hitting rate of the molecules to the receptors of an absorbing receiver in a 3D medium while varying the density and the size of receptors. Inspired by smart antennas in conventional wireless communications, the authors in \cite{felicetti2015smartAF} designed a specialized receiver to achieve directivity gain.  

In this letter, we analyze the propagation and reception pattern of the messenger molecules emitted from a spherical transmitter. Rather than assuming a point transmitter, we focus on a spherical transmitter to consider more realistic scenarios. Due to the reflecting spherical transmitter, directivity gain is obtained. We quantify the difference of power gain between the point and spherical transmitters. As achieving directivity gain is a crucial technique to boost the information rate in conventional wireless communications, directivity in molecular communication should also be analyzed and exploited. With this aim,  the authors introduced several metrics to specify directivity in molecular communication.

% % % % % % % % % % % % % % % % % % % % % % % % % % % % % % % %
% % % % % % % % % % % % % % % % % % % % % % % % % % % % % % % %
\section{System Model for Molecular Communication via Diffusion}
\label{sec_system_model}
\begin{figure}[t]
\begin{center}
	\includegraphics[width=1.0\columnwidth,keepaspectratio]%
	{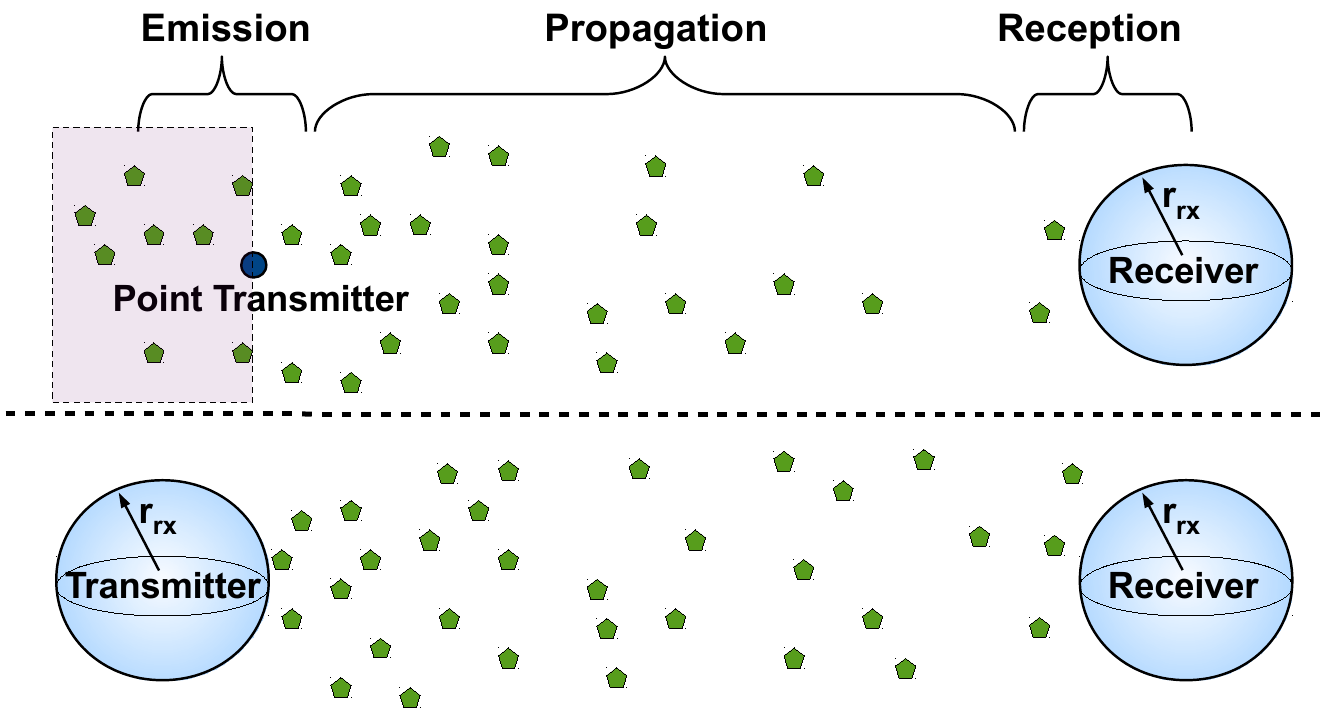}
	\caption{System model of MCvD with point and spherical transmitter cases. In the point transmitter case, molecules can go in the opposite direction, which is shown by a box covering such molecules. On the other hand, molecules are reflected by the body of the transmitter in the spherical transmitter case. }
	\label{fig_system_model}
\end{center}
\end{figure}

In an MCvD system, there is at least one transmitter and receiver pair in a fluid environment. In general, transmitter and receiver nodes are assumed to be point and spherical, respectively~\cite{kuran2012interferenceEO,yilmaz20143dChannelCF,mahfuz2014strengthBO,kim2013novelMT}. In this letter, we remove the assumption of having a \emph{point source} for the transmitter side. Instead, we consider a spherical transmitter which reflects the emitted molecules, i.e., the spherical transmitter obstructs the molecules that are trying to go in the opposite direction (Fig.~\ref{fig_system_model}). Therefore, the number of received molecules is expected to be higher, compared to the point source case.
% % % %

%In MCvD, information is modulated on some of the physical properties of the molecules. 
We assume that the receiver  is a perfectly absorbing node and whenever a molecule hits the surface of the receiver node, it contributes to the received signal. This type of process is called first passage process (FPP) and its hitting histogram exhibits inverse Gaussian distribution~\cite{redner2001guideTF,srinivas2012molecularCI}. This process model is also observed in most receptors (in biological systems) that react with the information molecules or that activate \emph{uptake mechanisms}~\cite{cuatrecasas1974membraneR}. Therefore, biological systems generally have mechanisms to assure single contribution by each molecule to the received signal. Considering FPP for the propagation and the reception is more realistic than considering the free diffusion with a passive receiver, though FPP leads to more challenges in the analytical derivations. 
%However, due to the realistic modeling concerns, we consider absorbing receiver for which the first passage process is the deriving model. 

In Fig.~\ref{fig_system_model}, point and spherical transmitter cases in a 3D environment are depicted. Emitted molecules diffuse in the environment which is characterized by diffusion coefficient~$D$. In both cases, we consider a perfectly absorbing receiver node which counts the number of absorbed molecules within a fixed duration. The received signal consists of the time histogram of hitting molecules. When we have point source,  molecules are able to go in the opposite direction of the receiver more freely. In the spherical transmitter case, however, molecules are obstructed and reflected by the transmitter. Hence, the received signals of these two cases are expected to differ.

%We assume that the receiver node can reset its counter and restart counting the hitting molecules in consecutive slots.

% % % % % % % % % % % % % % % % % % % % % % % % % % % % % % % %
% % % % % % % % % % % % % % % % % % % % % % % % % % % % % % % %
\section{Propagation Pattern for Chemical Signals}
\label{sec_propagation_pattern}
This section revisits the derivation of the received signal for the point source case. We then define and explain the measurement setup for the spherical transmitter case.

\subsection{Molecular Signal Emitted from Point Source}
The diffusion process basically models the average movement of particles in the concentration gradient. The derivative of the flux with respect to time results in Fick's Second Law in a 3D environment, given by
% % %
\begin{align}
\label{eqn_ficks3d}
\frac{\partial \prt}{\partial t} = D \nabla^2 \prt
\end{align}
where $\nabla^2$, $\prt$, and $D$ are the Laplacian operator, the molecule distribution function at time $t$ and distance $r$ given the initial distance $r_0$, and the diffusion constant. The initial condition is given by
% % %
\begin{align}
\label{eqn_initial_condition}
\prtzero = \frac{1}{4 \pi r_0^2} \delta(r - r_0),
\end{align}
and the boundary conditions by
% % %
\begin{align}
\label{eqn_bd_condition_one}
\lim_{r \rightarrow \infty} \prt = 0,
\end{align}
\begin{align}
\label{eqn_bd_condition_two}
D \frac{\partial \prt}{\partial r} = w \,\prt \text{, for } r = \rrn
\end{align}
where $\rrn$ and $w$ denote the radius of the receiver and the rate of reaction. The reaction rate with the receiver boundary is controlled by $w$.  Specifically, $w=0$ means a  nonreactive surface and, on the other hand, $w\rightarrow\infty$  corresponds to the boundary where every collision leads to an absorption. Also note that \eqref{eqn_bd_condition_one} reflects the assumption that the distribution of the molecules vanishes at a distance far greater than $r_0$.  

In~\cite{yilmaz20143dChannelCF}, the solution to this differential equation system is presented and analyzed from the perspective of channel characteristics. After finding the reaction rate, the authors presented the formula for the fraction of molecules that hit the receiver until time $t$, as follows:
% % %
\begin{align}
\begin{split}
F_\text{hit}^{3\text{D}} (t)=  \frac{\rrn}{d\!+\!\rrn} \,\text{erfc} \left( \frac{d}{\sqrt{4Dt\,}}\right) = \frac{2 \rrn}{d\!+\!\rrn} \,\Phi\left(\frac{-d}{\sqrt{2Dt}}\right)
\end{split}
\label{eqn_3d_frac_received_point_src}
\end{align}
where $d$, $\text{erfc}(.)$, and $\Phi(.)$ represent the distance, complementary error function, and the standard Gaussian cdf, respectively. Since the point source has circular symmetry, all the points at the same radius are equivalent in terms of number of received molecules. Due to the circular symmetry, the solution for the system of differential equations was enabled. For the spherical transmitter case, however, it is harder to derive the formulation of the number of received molecules.

\subsection{Molecular Signal Emitted from Spherical Source}
\begin{figure}[t]
\begin{center}
	\includegraphics[width=1.00\columnwidth,keepaspectratio]%
	{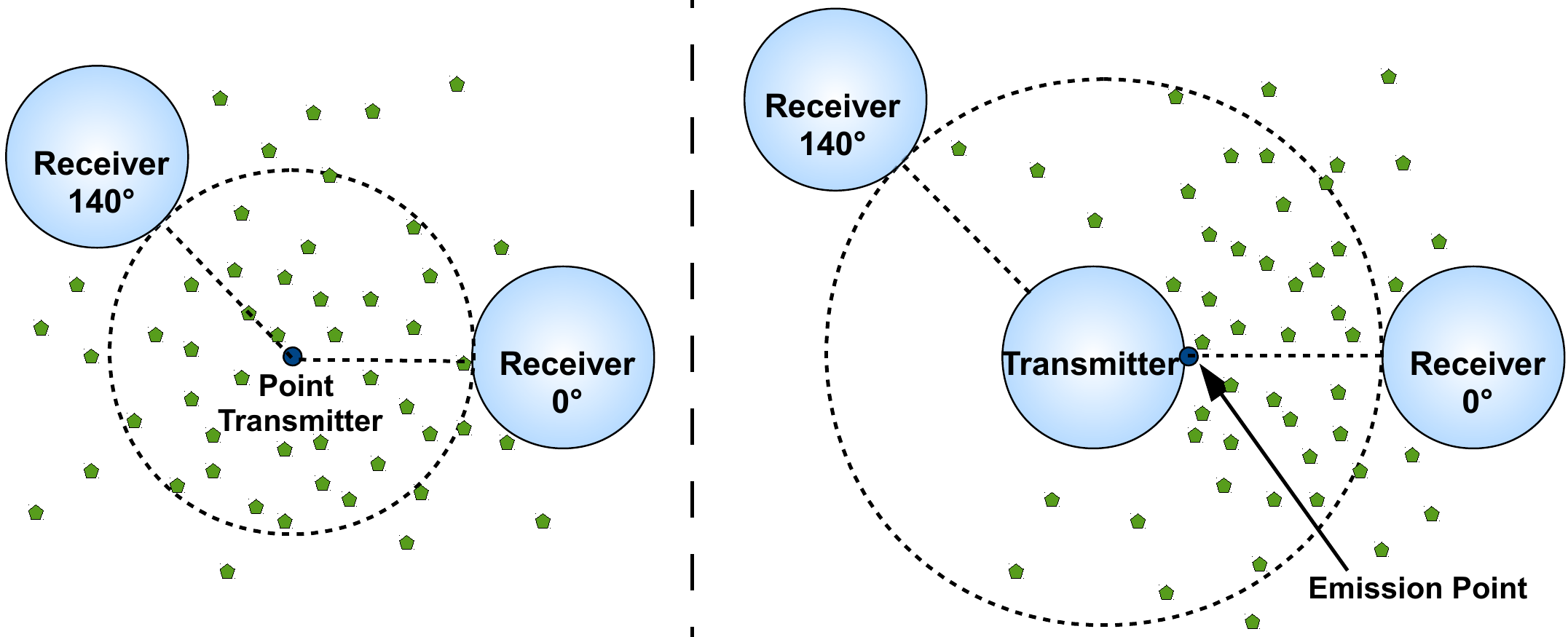}
	\caption{Measurement topology of point and spherical transmitter with the same distance and two example locations for the receiver node. }
	\label{fig_measurement_topology}
\end{center}
\end{figure}

For the spherical transmitter, we use a particle-based  simulator. We measure the directivity gain resulting from the non-symmetrical and obstructing body. In Fig.~\ref{fig_measurement_topology}, it can be easily seen that the spherical transmitter adds some directivity gain depending on the alignment of the receiver (e.g., if the receiver is on the back side of the emission point, it becomes harder for the molecular signal to reach the receiver). For the point source, on the other hand, equidistant points are equivalent and we use the analytical formulation given in~\eqref{eqn_3d_frac_received_point_src} to find the received signal at each angle. In Fig.~\ref{fig_measurement_topology}, the measurement topology is shown and we use \SI{10}{\degree} for the angle step while doing the \emph{molecular signal}  measurements.  
% % % %

% % % % % % % % % % % % % % % % % % % % % % % % % % % % % % % %
% % % % % % % % % % % % % % % % % % % % % % % % % % % % % % % %
\section{Results}
\label{sec_results}
\begin{figure*}[t]
\begin{center}
	\subfigure[$\rtn = \SI{2.5}{\micro\metre}$]
	{\includegraphics[width=0.67\columnwidth,keepaspectratio]
	{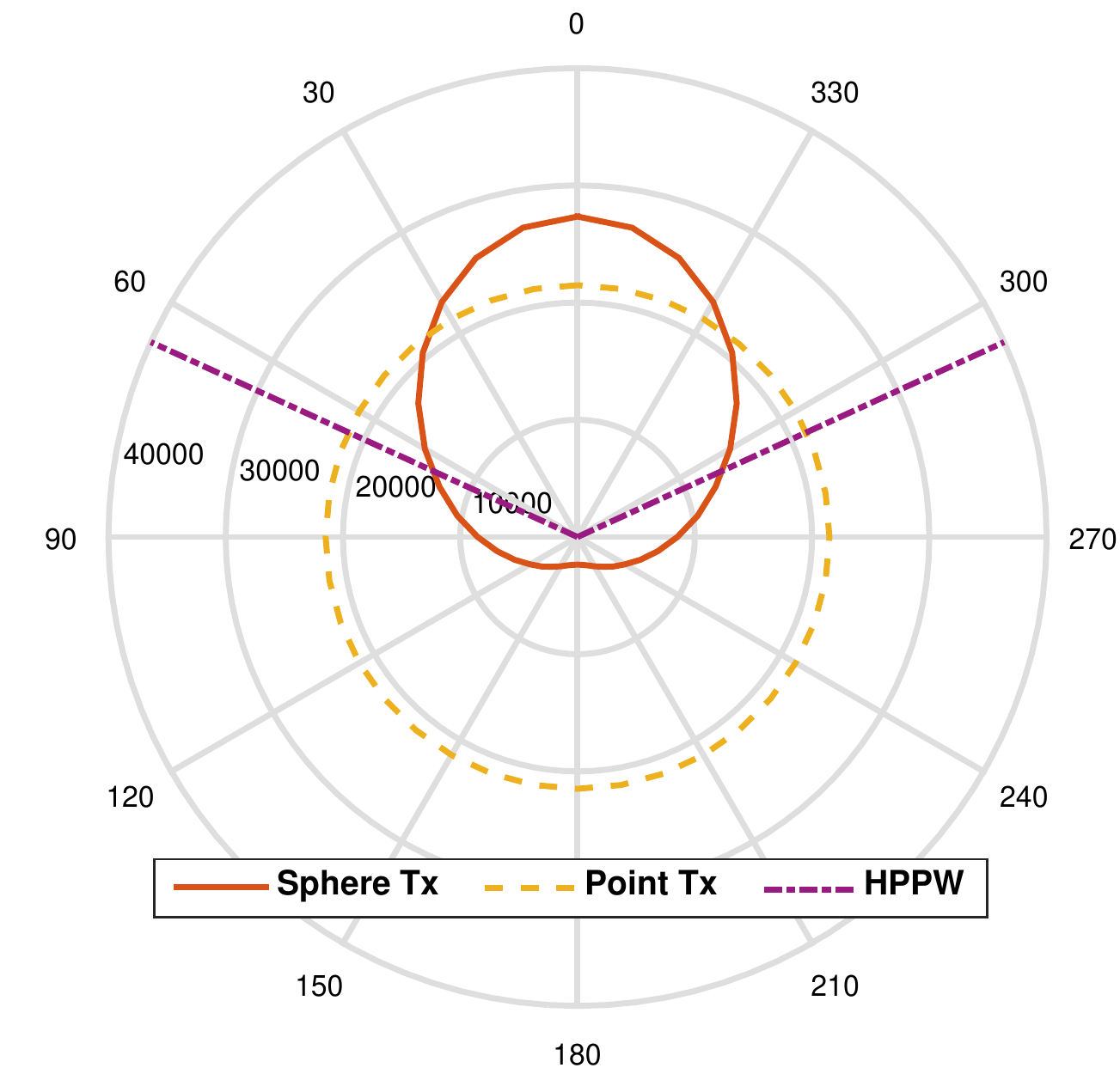}%
	\label{fig_polar_plot_25}}
	%\hfil
	\subfigure[$\rtn = \SI{5.0}{\micro\metre}$]
	{\includegraphics[width=0.67\columnwidth,keepaspectratio]
	{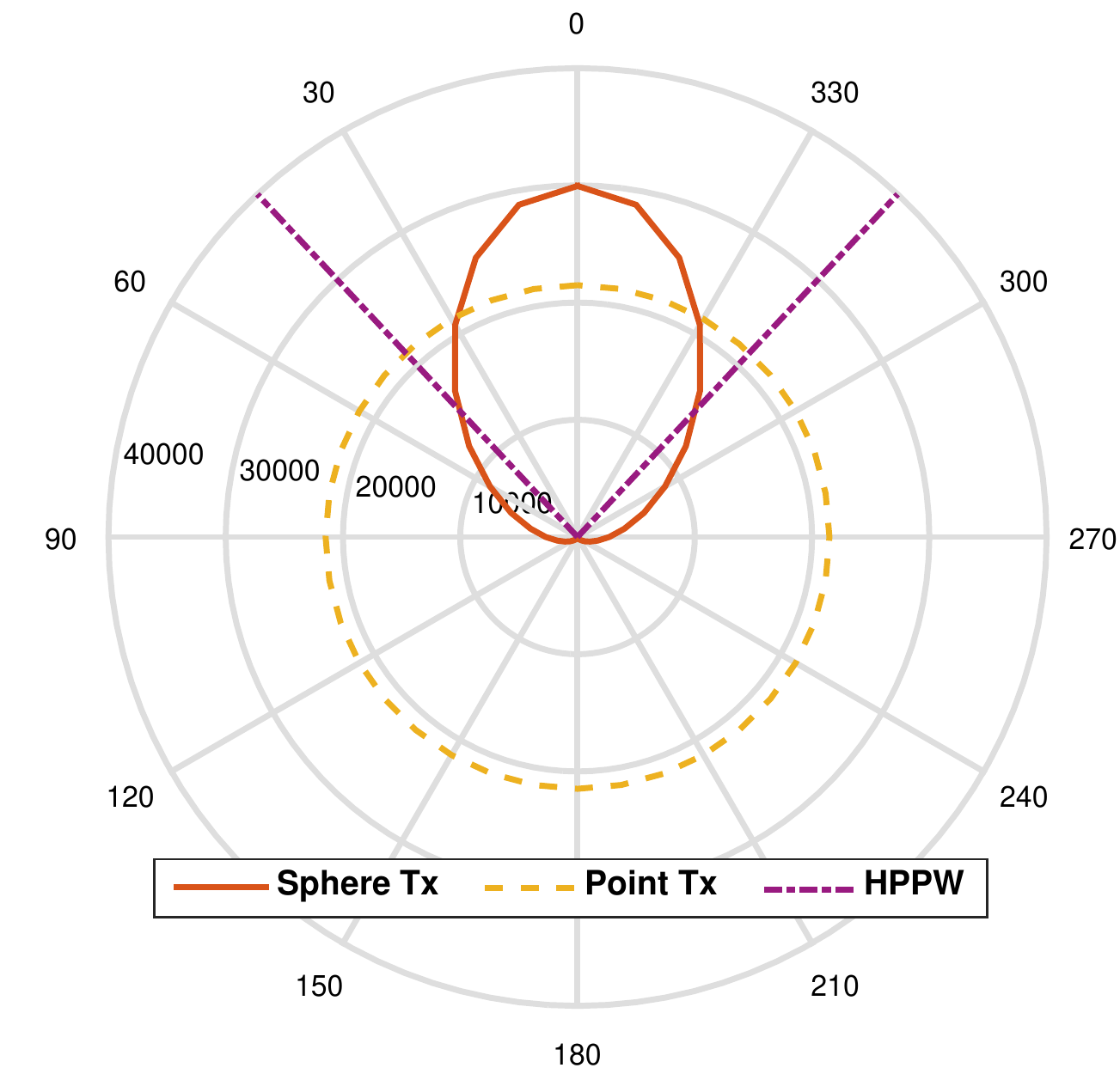}%
	\label{fig_polar_plot_50}}
	%\hfil
	\subfigure[$\rtn = \SI{7.5}{\micro\metre}$]
	{\includegraphics[width=0.67\columnwidth,keepaspectratio]
	{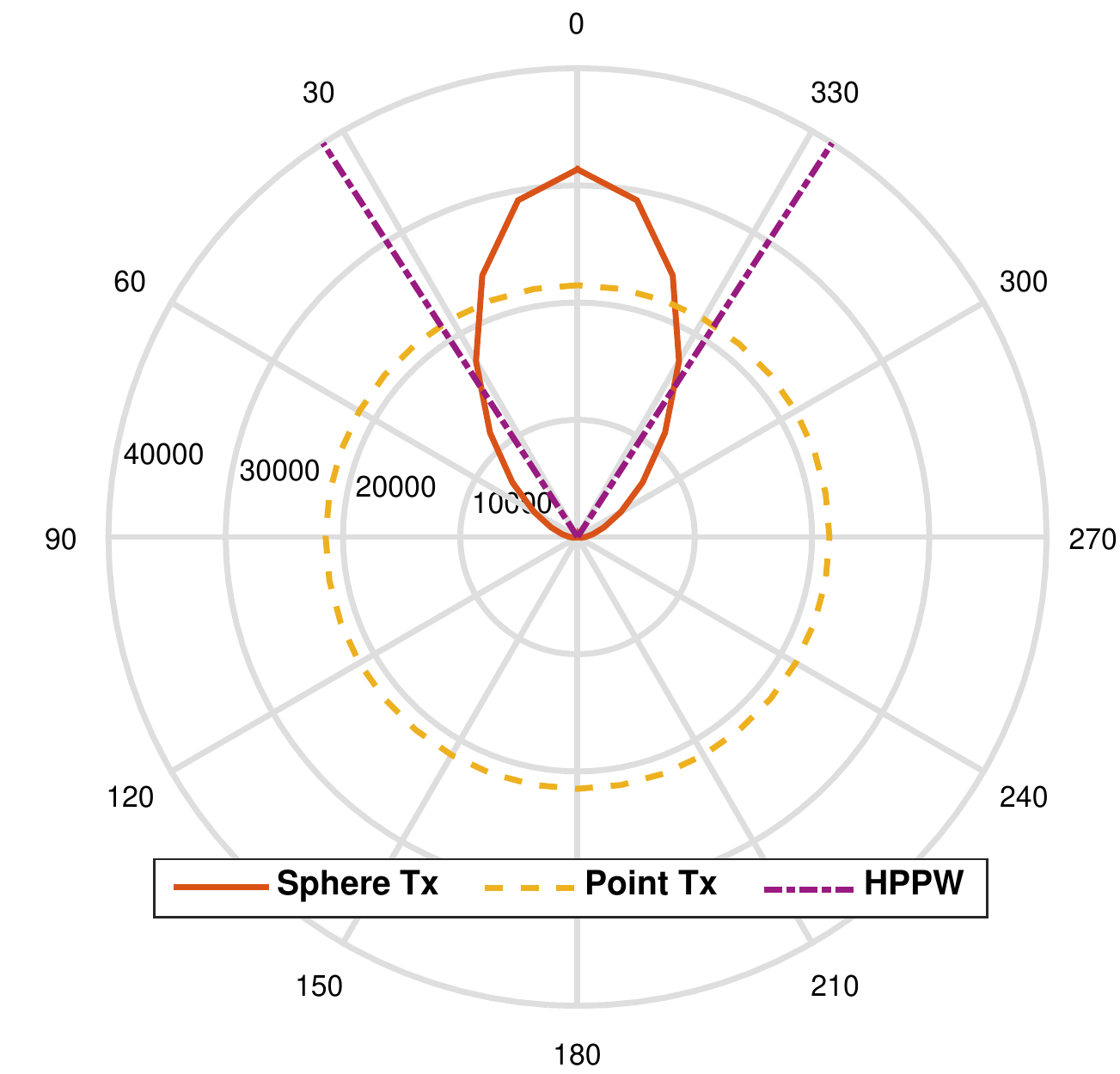}%
	\label{fig_polar_plot_75}}
	\caption{Polar plot of propagation patterns with different $\rtn$ values  ($d=\SI{2}{\micro\metre}$, $t_s=\SI{0.2}{\second}$). Received power is the same for the point transmitter case at each angle. In the spherical transmitter case, depending on the receiver location (angle) the received power is plotted (receiver antenna pattern).}
	\label{fig_polar_plot}
\end{center}
\end{figure*} 

\subsection{Parameters}
For the performance evaluation, we fixed some parameters and observed the effect of others. First, we give the polar plot of the chemical signal reception power pattern for the spherical transmitter and compare with that of the point transmitter case. We present common system parameters in Table~\ref{tbl_system_parameters}.

\subsection{Analysis Metrics}
We mainly focus on three metrics: half-power pattern-width (HPPW), directivity gain at each angle with respect to \emph{point source}, and peak time of the signal at each angle. Due to half symmetry in the spherical transmitter case, we get measurements of these metrics for the angles between $\SI{0}{\degree}$ and $\SI{180}{\degree}$, where $\SI{0}{\degree}$ corresponds to the perfect alignment of the transmitter and receiver nodes. We denote the number of received molecules at the angle $\alpha$ for the spherical transmitter until time $t$ as $\nrxTsph{\alpha}{t}$. For HPPW, we solve the following equation for $\alpha$  
% % %
\begin{align}
\nrxTsph{\alpha}{t_s} = \frac{1}{2} \nrxTsph{0}{t_s}\, , \;\; \text{HPPW}=2\alpha
\end{align}
where $t_s$ corresponds to the symbol duration (we generally consider this duration for analysis metrics). Smaller HPPW is a desired thing to have higher directivity gain.  

For the second metric, we measure the peak time of the received signal at each angle $\alpha$ and it is denoted by $\peakAt{\alpha}$.  Peak time determines the signal propagation delay. 

For the third metric, we normalize the number of received molecules at each angle with the case of the point source. As noted before, the number of received molecules does not change with the angle for the point source case (due to perfect circular symmetry). Therefore, we can use any angle for the point source scenario that corresponds to the normalization value. Hence, we formulate the directivity gain at angle $\alpha$ as follows:
% % %
\begin{align}
\directivityG{\alpha} = \frac{\nrxTsph{\alpha}{t_s}}{\nrxTpnt{\alpha}{t_s}}   \, .
\end{align}

\begin{table}[!t]
\begin{center}
\caption{Range of parameters used in the experiments}
\renewcommand{\arraystretch}{1.14}
\label{tbl_system_parameters}
\begin{tabular}{p{5.2cm} l}
\hline
\bfseries{Parameter} 							& \bfseries{Value} \\ 
\hline 
Number of emitted molecules			& $40\,000$\\
Diffusion coefficient ($D$) 		& $\SI{100}{\micro\metre^2/\second}$\\
Distance ($d$) 						& $\{2,\, 4,\, 6\}\,\, \si{\micro\metre} $\\
Transmitter radius ($\rtn$)			& $\{0,\, 2.5,\, 5,\, 7.5\}\,\,\si{\micro\metre}$\\
Receiver radius ($\rrn$)			& $\SI{5}{\micro\metre}$\\
Simulation duration ($\tend$) 		& $d^2 \, \SI{0.1}{\second/\micro\metre^2}$\\
\hline
\end{tabular} 
\end{center}
\end{table}
\subsection{Polar Plot of Propagation Pattern}
In Fig.~\ref{fig_polar_plot}, we plot $\nrxTsph{\alpha}{t_s}$ and $\nrxTpnt{\alpha}{t_s}$ values where the latter one does not change with the angle (due to circular symmetry). We also plot the HPPW as a reference. The first observation is that when we increase the transmitter node size the HPPW narrows. Also note that when we increase $\rtn$, the received power increases gradually due to the slight guidance given by the transmitter node.

It can be clearly seen that having a spherical reflecting transmitter affects the received power. We can also claim that, for the spherical transmitter case, if the receiver node is placed  between $\SI{-20}{\degree}$ and $\SI{20}{\degree}$ (i.e., if the receiver and the transmitter are well aligned), then the received power is higher for the given parameters. On the other hand, if we consider the reception on the back side of the transmitter, we observe that received power diminishes quickly. In the case of a well-aligned receiver, we can consider these low power values as the interference leakage to other receivers on the back side of the transmitter.

\subsection{$\peakAt{\alpha}$ Analysis}

\begin{figure}[t]
\begin{center}
	\includegraphics[width=0.97\columnwidth,keepaspectratio]%
	{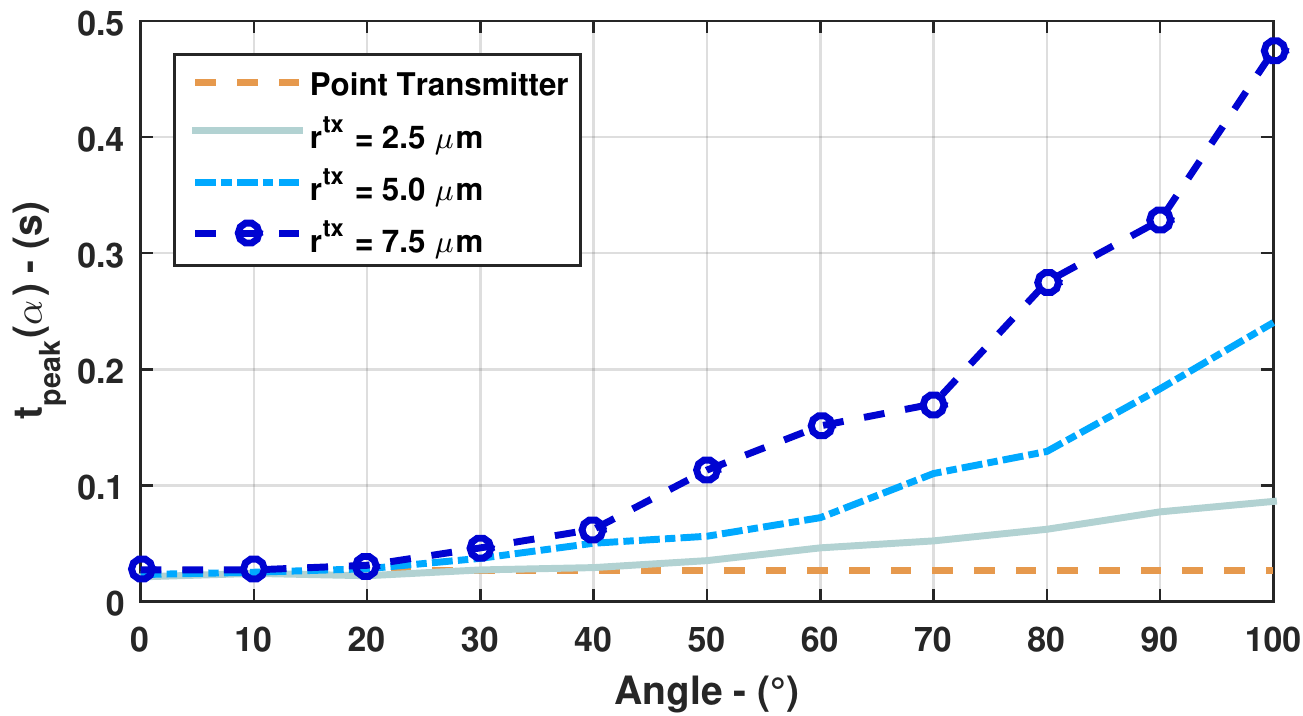}
	\caption{Angle versus peak time of the received signal ($d=\SI{4}{\micro\metre}$, $t_s=\SI{0.8}{\second}$). }
	\label{fig_tpeak_d4}
\end{center}
\end{figure}

In Fig.~\ref{fig_tpeak_d4}, we present $\peakAt{\alpha}$ values for different angles in the measurement topology. For the point transmitter case, $\peakAt{\alpha}$ does not change with the angle due to circular symmetry. For the spherical transmitter case, as we place the receiver at wider angles, the received signal peak time also increases (i.e., the chemical signal reaches with more delay). Also note that the relation between angle and $\peakAt{\alpha}$ is not linear, which is possibly due to the derived relation $t_{\text{peak}} \varpropto d^2$ in~\cite{yilmaz20143dChannelCF} for the point transmitter case. For the small angles, although $\peakAt{\alpha}$ values are similar, the difference becomes notable at the wider angles. If we place the receiver at wider angles and have a larger transmitter,  we observe a higher delay for the received signal. If we also consider Fig.~\ref{fig_polar_plot}, then we can claim that larger transmitters experience more delay and lesser power leakage at the back side of the transmitter. 
% % % %

\subsection{Directivity Gain \& HPPW Analysis}
% % % %
\begin{figure}[t]
\begin{center}
	\includegraphics[width=0.97\columnwidth,keepaspectratio]%
	{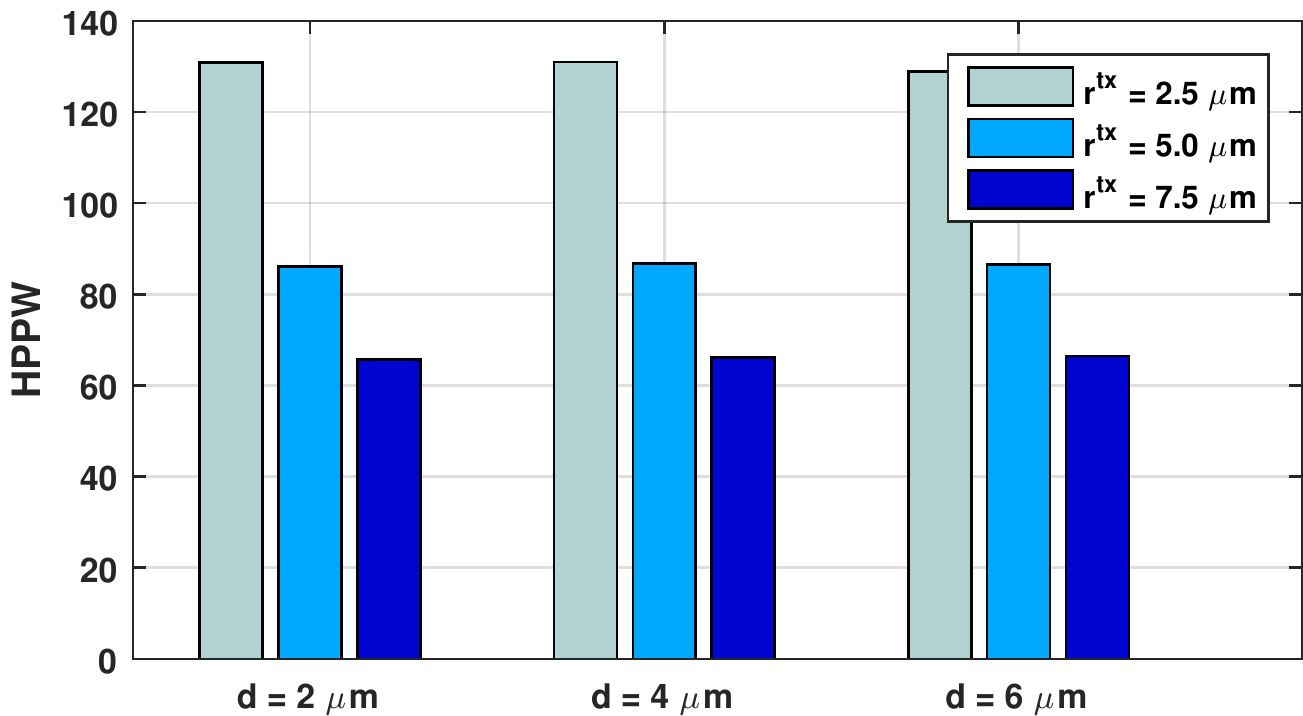}
	\caption{Half-power pattern-width plot ($t_s=\SI{0.2}{\second}$ for all distances).}
	\label{fig_hpbw}
\end{center}
\end{figure}
In Fig.~\ref{fig_hpbw}, the half-power pattern-width values for different distances and $\rtn$ values are depicted. We fix  $t_s$ and obtain the measurements that lead us to derive the HPPW values. When we increase $\rtn$, HPPW value decreases and the propagation and reception of chemical signal becomes more directed and narrower. In Fig.~\ref{fig_polar_plot}, we also see a narrower pattern for larger transmitter cases. On the other hand, the effect of distance is nearly negligible for the fixed $t_s$ value. 
% % % %
\begin{figure}[t]
\begin{center}
	\includegraphics[width=0.97\columnwidth,keepaspectratio]%
	{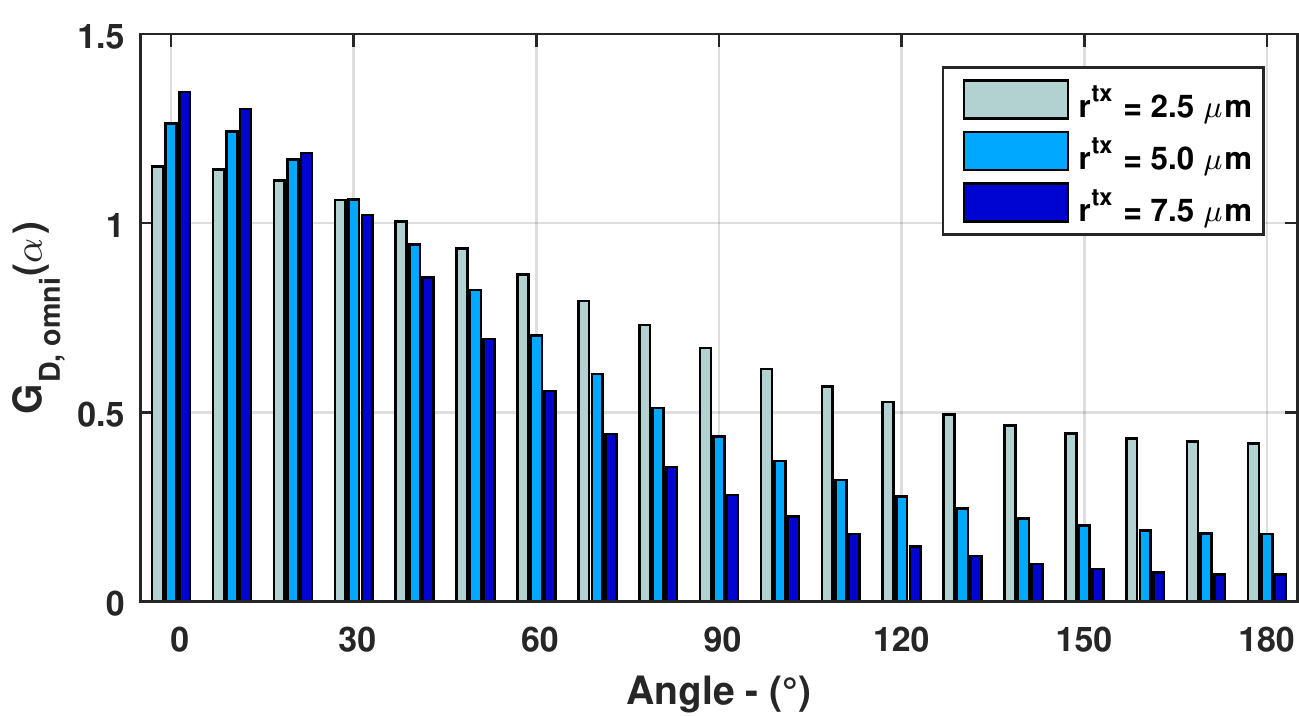}
	%\caption{Angle versus directivity gain with respect to point transmitter case ($d=\SI{6}{\micro\metre}$, $t_s=\SI{1.8}{\second}$).}
	\caption{Angle versus directivity gain ($d=\SI{6}{\micro\metre}$, $t_s=\SI{1.8}{\second}$).}
	\label{fig_directivty_gain}
\end{center}
\end{figure}
In Fig.~\ref{fig_directivty_gain}, the directivity gain with respect to point transmitter is plotted for different angles. For the well-aligned receiver cases (i.e., $\alpha \leq 20$), $\rtn=\SI{7.5}{\micro\metre}$  has a higher gain. For the miss-aligned receiver cases, the gain decreases faster for  $\rtn=\SI{7.5}{\micro\metre}$  compared to the other cases. Therefore, the $\SI{7.5}{\micro\metre}$ case creates less \emph{molecular signal leakage} for the other receivers at the wider angles.

% % % % % % % % % % % % % % % % % % % % % % % % % % % % % % % %
% % % % % % % % % % % % % % % % % % % % % % % % % % % % % % % %
\section{Conclusion}
In this study, we have analyzed the received signal for the point and spherical transmitter cases when the receiver is an absorbing receiver. If the transmitter is a reflecting spherical body, then a change occurs in the molecule propagation and the average distribution of molecules in the environment. To understand the propagation and reception pattern, we conducted simulations by changing the receiver location on the same radius for directivity analysis. We analyzed the propagation and reception pattern in terms of new metrics in MCvD such as $\peakAt{\alpha}$, directivity gain, and HPPW. Our analysis showed that a larger transmitter radius yields a narrower reception pattern. That is if the receiver and the transmitter are well-aligned then the received power increases. On the other hand, reception at the wider angles (or possibly on the back side of the transmitter) is weak, which can be interpreted as power leakage to the other receivers. We also quantified the HPPW for the equal duration while changing the distance and the transmitter radius. For future work, we plan to design and analyze a communication system utilizing this concept~\cite{koo2016molecularMimo_jsac}.
%derive the analytical formulation of the received signal and characterize the channel. 

%\section*{Acknowledgment}

\bibliographystyle{IEEEtran}
\bibliography{antennaMC_yearSorted}

\end{document}